\documentclass[titlepage,12pt]{article}  
\usepackage{amsfonts}  
\usepackage{psfig}  

\newcommand{\R}{{\mathbb R}}

\newcommand{\be}{\begin{equation}} 
\newcommand{\ee}{\end{equation}} 
\newcommand{\bea}{\begin{eqnarray}} 
\newcommand{\eea}{\end{eqnarray}}

\newcommand{\cQ}{{\mathcal{Q}}} 
 
\newcommand{\cF}{{\mathcal{F}}} 
\newcommand{\cG}{{\mathcal{G}}}

\newcommand{\cH}{{\mathcal{H}}}

\newcommand{\cA}{{\mathcal{A}}}

\newcommand{\nn}{\nonumber}

\begin{document} 
\title{Complex Instantons and Charged Rotating Black Hole Pair Creation}  
\author{I. S. Booth and R. B. Mann\\         
             Department of Physics\\         
             University of Waterloo\\         
             Waterloo, Ontario\\         
             N2L 3G1} 
 \date{June 4, 1998\\} 
\maketitle\begin{abstract} 
We consider the general process of pair-creation of charged rotating black  
holes.  We find that instantons which describe this process are necessarily  
complex due to regularity requirements. However their associated  
probabilities are real, and fully consistent with the interpretation that the  
entropy of a charged rotating black hole is the logarithm of the number of  
its quantum states. 
\end{abstract}  
\pagenumbering{arabic}

Black hole pair production is by now a well-established process in   
semi-classical  quantum gravity.  Pair-production   
is a tunnelling process in which the mass-energy of the   
created pair of black holes is balanced by their negative potential energy  
in some  background field. Studies of this phenomenon for non-rotating black holes
have repeatedly   
provided us with evidence that the exponential of the entropy of a black hole does   
indeed correspond to the number of its quantum states.  It is quite robust, with   
many different background fields -- an electromagnetic field  
\cite{dgkt}, a positive cosmological constant \cite{cosern},  
a cosmic string \cite{strpair}, or a domain wall \cite{dompair} -- yielding  
qualitatively similar features for the pair production process.  
The probability amplitude for the process is approximated by $e^{-I_i}$,   
where $I_i$ is the action of the relevant instanton {\it i.e.} a Euclidean  
solution to the field equations which interpolates between the states  
before and after a pair of black holes is produced.    
  
In this paper we consider this problem for the rotating case
and show that we must necessarily modify the definition of  
instanton beyond that which is given by the usual periodic-imaginary  
time formalism \cite{OrigPathInt}. This formalism, originally  
developed to enable the computation of tunnelling processes and canonical  
partition functions \cite{Hibbs} has (apparently) led to the belief  
that the relevant instanton action must always be obtained from real,  
positive-definite ({\it i.e.} Euclidean)  metrics.  For rotating black  
holes, the real Euclidean metric is then obtained by supplementing  
the analytic continuation $t\to it$ with the transformation $J\to  
iJ$, where $J$ is the real angular momentum \cite{GibH}. We find that  
this prescription cannot consistently define an instanton that will  
mediate the creation of a pair of charged rotating black holes in a  
background with a positive cosmological constant, and that 
complex instantons must be employed to describe the  
pair-production of rotating black holes. To our knowledge this is the
first example in which naive analytic continuation of parameters in a 
stationary
black hole metric will not correctly describe the relevant physical process.

The relevance of complex  
metrics for black hole thermodynamics was discussed in ref.\  
\cite{BMY}, and our results are consistent with this interpretation  
of the functional integral formalism. Although  
there is no real Euclidean instanton that describes the pair  
production of rotating black holes, the action associated with the complex  
instanton is real, and describes the probability of the pair creation  
of rotating black holes in a manner fully consistent with the  
interpretation of the entropy as the logarithm of the number of  
quantum states of the black hole, as in the non-rotating case.

To illustrate this we shall take our background energy to be that
provided by a cosmological constant, and consider two oppositely 
charged black holes undergoing uniform acceleration in this background.
This physical situation is described by the following metric \cite{PlebDem}
%The $C$ metric solution of the Einstein-Maxwell equations  
%may be interpreted as describing two oppositely-charged black holes  
%undergoing uniform acceleration \cite{KW}. Its generalization  
%to include charged and rotating black holes in a cosmological background  
%is \cite{PlebDem},   
\begin{equation}   
\label{PlebDemMetric}   
ds^2  = \frac{1}{(p-q)^2} \left[ \frac{1+p^2q^2}{P} dp^2    
+ \frac{P}{1+p^2q^2} \left( d \sigma - q^2 d \tau \right)^2    
- \frac{1+p^2q^2}{Q} d q^2    
+ \frac{Q}{1+p^2q^2} \left(d \tau  + p^2 d \sigma \right)^2   
\right],   
\end{equation}
generalizing the $C$ metric which describes the
spacetime associated with a pair of uniformly accelerating neutral spinless 
black holes \cite{KW}. The accompanying electromagnetic 
field is defined by the vector potential  
\be  
\label{PDA}  
A = - \frac{e_0 q (d \tau + p^2 d\sigma)}{1 + p^2 q^2} + \frac{g_0 p (d   
\sigma - q^2 d\tau)}{1 + p^2 q^2},  
\ee  
where $p,q,\tau$, and $\sigma$ 
(which is periodically identified with some period $T$) are coordinate functions,   
$P(p) = (-\frac{\Lambda}{6} - g_0^2 + \gamma)  - \epsilon    
p^2    
+ 2 m p^3 + (- \frac{\Lambda}{6} - e_0^2 - \gamma) p^4$,   
and $Q(q) = P(q) + \frac{\Lambda}{3} (1 + q^4)$. $\Lambda$ is the    
cosmological constant (assumed positive),    
$\gamma$, $\epsilon$ are constants connected    
in a non-trivial way with rotation and acceleration, $e_0$ and $g_0$    
are linear multiples of electric and magnetic charge, and $m$ is   
the mass parameter. This solution may be analytically   
extended across the singularity at $p=q$. Then the  
two accelerating black holes are on opposite sides of 
that  $p=q$ hypersurface. 
    
Instantons associated with the metric (\ref{PlebDemMetric}) can
then be interpreted as describing the pair-creation of rotating charged
black holes in a cosmological vacuum provided that the periodic identification
of $\sigma$ does not force the inclusion of additional sources of stress-energy (e.g. struts).
%Instantons associated with the metric (\ref{PlebDemMetric}) can
%then be interpreted as describing the pair-creation of rotating charged
%black holes in a cosmological vacuum, provided that no additional sources
%of stress-energy (e.g. struts) are required. 
These are required if there are conical singularities 
in (\ref{PlebDemMetric}). By matching the accelerations of the holes to 
the source of background energy, such conical singularities 
(located at the roots of $P$) may be eliminated from (\ref{PlebDemMetric})
as follows.
%The space-time described by (\ref{PlebDemMetric}) contains conical singularities,
%which (without introducing additional struts of stress-energy) 
%must be eliminated by matching the accelerations of the holes to the 
%cosmological background energy. 
%Periodically identifying $\sigma$, possible 
%Potential conical    
%singularities are located at the roots of $P$ and may be eliminated as
%follows. 
Suppose $P$ has two real roots at $p_0\pm\alpha$ and two complex roots at  
$\hat{p} \pm i \beta$, where $p_0, \alpha, \hat{p}, \beta \in \R$.
If $P(p)$ has an axis of symmetry along the line $\hat{p}=p_0$
then conical singularities at $p_0\pm\alpha$ may be
simultaneously eliminated by identifying $\sigma$ with period $T =    
\frac{4 \pi}{P'(p_0-\alpha)}$, where $P'=\frac{dP}{dp}$. Making a series  
of $\beta$ dependent parameter rescalings and coordinate transformations   
(which up to linear factors associate $p \leftrightarrow p_0 + \alpha \cos   
\theta$, $q \leftrightarrow \frac{1}{r}$, $\sigma \leftrightarrow \phi$, and   
$\tau \leftrightarrow t$), and then  taking the $\beta \rightarrow 0$ limit, we   
obtain the well-known Kerr-Newmann-deSitter (KNDS) solution  
\cite{MM}:  
\bea   
\label{KNdS}   
ds^2  =  -\frac{\cQ}{\cG \chi^4} \left(  dt - a \sin^2\theta d\phi   
\right)^2 + \frac{\cG}{\cQ} dr^2   
      + \frac{\cG}{\cH} d\theta^2 + \frac{\cH \sin^2 \theta}{\cG \chi^4}  
\left( a dt - \left[ r^2 + a^2 \right] d\phi \right)^2,   
\eea   
where   
\bea  
&& \cG \equiv r^2 + a^2 \cos^2 \theta, \; \; \;  \cH = 1 +  
\frac{\Lambda}{3} a^2 \cos^2 \theta, \; \; \; \chi^2 = 1 +  
\frac{\Lambda}{3} a^2, \mbox{ and} \nn \\   
&&\cQ = -\frac{\Lambda}{3} r^4 +  
\left( 1 - \frac{\Lambda}{3} a^2 \right) r^2 - 2Mr + \left( a^2 + E_0^2 +  
G_0^2 \right).  \nn   
\eea   
and where
\be  
\label{EMfield}   
F = -\frac{1}{\cG^2 \chi^2} \left\{ X dr \wedge (dt - a  
\sin^2 \theta d\phi) + Y \sin \theta d\theta \wedge ( a dt -  
(r^2+a^2)d\phi) \right\},   
\ee   
is the electromagnetic field, with $X = E_0 \Gamma + 2aG_0 r \cos \theta  
$, $Y = G_0 \Gamma - 2 a E_0 r \cos \theta$, and $ \Gamma = r^2 - a^2  
\cos^2 \theta$, $M$ is the black hole mass, $a$ is the rotation  
parameter, and $E_0$ and $G_0$ the respective electric and magnetic charges. 

Hence cosmological pair creation of charged rotating black  
holes reduces to a consideration of the nonsingular instantons that can be  
constructed from the KNDS metric. We have checked that most of the  
other methods of eliminating conical singularities from the metric
(\ref{PlebDemMetric}) also lead to this class  
(although a few yield a broader class of metrics), and so we shall only  
consider KNDS instantons in the sequel.  
  
We consider cases where the polynomial $\cQ$ has four real roots,  
three of which are positive and correspond (in decreasing order) to  
the cosmological, outer and inner black hole horizons.
Under appropriate periodic identifications, the maximal extension
yields a space-time whose $t= \mbox{constant}$ hypersurfaces 
each contain only a single pair of black holes with equal but opposite  
charge and rotation.  
To construct regular instantons which mediate rotating black hole
pair-creation, we foliate space-time with a set of  
space-like hypersurfaces $\Sigma_t$ labeled by a time coordinate $t$   
(whose forward pointing unit normal is $u_a = - N dt$), so that the metric is
\be  
\label{general} ds^2 = - N^2 dt^2 + h_{ij} (dx^i + V^i d t) (dx^j + V^j  
dt),   
\ee   
where $h_{ij}$ is the induced metric on the hypersurfaces, $N$ is  
the lapse function, and $V^i$ is the shift vector field (a three vector  
field defined on each hypersurface). The extrinsic curvature  
$K_{ij}$ and the 3-vectors $E^i$ and $B^j$ that describe the electric and  
magnetic fields defined on $\Sigma$ must satisfy the constraints 
\bea  
\cH &\equiv& ^{(3)}R + K^2 - K^{ij}K_{ij} - 2 (E^2 + B^2) = 0  
\label{MEC1}\\  
\cH_i &\equiv& D_j K_i^{\; j} - D_i K - 2 \varepsilon_{ijk}E^j B^k = 0  
\label{MEC2}\\       
\cF_{el} &\equiv& D_j E^j = 0 \label{MC1}\\  
\cF_{mag} &\equiv& D_j B^j = 0 \label{MC2},  
\eea  
found by projecting the Einstein-Maxwell equations into the hypersurface,   
so that the specified field configuration is one admitted by the full  
field equations. Here $^{(3)}R$ is the Ricci scalar   
for $(\Sigma, h)$, $K = h^{ij}K_{ij}$,  
$E^2 = h_{ij}E^iE^j$, $B^2 = h_{ij}B^iB^j$, $D_j$ is the covariant  
derivative on $\Sigma$ that is compatible with $h_{jk}$, and  
$\varepsilon_{ijk}$ is a Levi-Cevita tensor.  
Given a full solution $\{M,g_{\alpha \beta},F_{\alpha \beta}\}$  
to the equations of motion and a surface $\Sigma$, the surface fields  
$h_{ij}$, $K_{ij}$, $E_i$, and $B_i$ on $\Sigma$ are induced via the  
relations $h_{ij} = e_i^\alpha e_j^\beta g_{\alpha \beta}$, $K_{ij}  
= - e_i^\alpha e_j^\beta \nabla_\alpha u_\beta$, $E_i = e_i^\alpha  
F_{\alpha \beta} u^\beta$, and $B^i = - \frac{1}{2} e^i_\alpha  
\varepsilon^{\alpha \beta \gamma \delta} u_\beta F_{\gamma \delta}$  
where the $e^i_\alpha / e_i^\alpha$ are the projection  
operators from the tangent/cotangent spaces to points  
in $M$ (that are also in $\Sigma$) into the  
intrinsic tangent/cotangent spaces of $\Sigma$.    
  
To find the probability amplitude for black hole pair-creation
we must match the instanton smoothly onto a Lorentzian solution 
describing the charged rotating pair of black holes across   
the hypersurface $\Sigma$, which implies that   
the induced metric field $h_{ij}$ and extrinsic curvature $K_{ij}$   
must be continuous there, along with the induced electric and magnetic  
fields. Using $\{x^1=\phi,x^2=\theta,x^3=r\}$ as coordinates on $\Sigma$   
we obtain  
\bea \label{matchup}  
h_{ij} &\equiv& e_i^\alpha e_j^\beta (g_{\alpha \beta} + u_\alpha u_\beta)  
 =  
\mbox{diag}[h_{\phi \phi}, h_{\theta \theta}, h_{rr}],   
\\ K_{ij} &\equiv&  
e_i^\alpha e_j^\beta u_{\alpha;\beta} = \left[  
         \begin{array}{ccc}   
    0 & \frac{h_{\phi \phi} \partial_\theta V^\phi}{2N} &   
        \frac{h_{\phi \phi} \partial_r V^\phi}{2N} \\   
    \frac{h_{\phi \phi} \partial_\theta V^\phi}{2N} & 0 & 0 \\   
    \frac{h_{\phi \phi} \partial_r V^\phi}{2N} & 0 & 0   
         \end{array}   
         \right], \\ E_i &\equiv& e_i^\alpha F_{\alpha \beta} u{^\beta} =  
\left[0 , \frac{F_{\theta t} - F_{\theta \phi} V^\phi}{N}, \frac{F_{r t} -  
F_{r \phi}V^\phi}{N}\right], \mbox{ and} \\ B_i &\equiv& -\frac{1}{2}  
e_i^\alpha g_{\alpha \beta} \varepsilon^{\beta \gamma \delta \gamma}u_c  
F_{\delta \epsilon} = \left[0, -\frac{h_{\theta \theta} F_{\phi  
r}}{\sqrt{h_{\phi \phi}h_{\theta \theta} h_{rr}}}, \frac{h_{rr} F_{\phi  
\theta}}{\sqrt{h_{\phi \phi}h_{\theta \theta} h_{rr}}} \right]  
\eea   
for these quantities.
  
To construct an instanton that can match onto $\Sigma_t$ and these  
fields, we map $N \rightarrow i N$, $V^j \rightarrow i  
V^j$, $F_{t j} \rightarrow iF_{t j}$, $F_{j t} \rightarrow i F_{j t}$ (for  
$j \in \{ \phi, \theta, r\}$), which is equivalent to sending $t\to i\tau$. 
This map preserves the matching quantities  
$h_{ij}$, $K_{ij}$, $E_i$, and $B_i$, the constraint  
equations (\ref{MEC1} -- \ref{MC2}) and the dynamical equations of   
motion \cite{BMY}.  We then identify the $\tau$ coordinate with some  
period $T$ and further identify this ``time"  coordinate as a single  
time at the horizons. Ensuring regularity is a little more  
difficult given that the metric as a whole is now complex (easily seen
by setting $t\to i\tau$ in (\ref{KNdS})), but if we  
examine the hypersurfaces of the instanton where $\theta = 0$ and  
$\theta = \pi$ we may again work with a real two dimensional Euclidean 
metric on the surface $\theta=0$ (or $\pi$), and eliminate 
conical singularities by making an appropriate  
choice of values for the physical parameters $\Lambda$, $M$,  
$a$, $E_0$, and $G_0$ (for a full description see ref. \cite{me}).   
  
The resultant complex solutions may be sliced in half by making cuts  
along the $\tau =0$ and $\tau = \frac{T}{2}$ hypersurfaces. This gives
a hemisphere which may then be attached to a Lorentzian KNdS space-time  
along a $t = \mbox{constant}$ hypersurface without difficulty, providing   
us with an instanton that describes the creation of a  
space-time containing two rotating charged black holes. 
Several admissible classes of instantons arise (generalizing earlier 
cases \cite{cosern}) which are categorized by their relative black
hole and cosmological horizon temperatures. We refer to the class in
which both
temperatures are equal as 
the lukewarm class. Here the $t = \mbox{constant}$
spatial sections consist of two non-degenerate black holes in thermal equilibrium
on opposite sides of the cosmological horizon; the analogous case where 
the black hole horizons are degenerate
is called the cold class. The non-degenerate limit
in which the cosmological and outer black hole horizons coalesce is called the 
charged-rotating Nariai class.
Although it does not contain any black holes,   
for non-rotating cases the Nariai case is unstable   
with respect to quantum tunelling into a more  
standard RNdS space-time \cite{Nar}, and so is relevant to  
consider. The classes in which all three horizons are coincident are referred to as
ultracold; it is not clear that these space-times have anything to do with black  
hole pair creation.
  
To compute the probability of creating a pair of black holes within one of 
these classes we must carry out a proper calculation of the associated
instanton action. This requires the  
appropriate inclusion of boundary terms, which are necessary to  
hold physical quantities constant so that the boundary data  
(\ref{matchup}--13) on $\Sigma$ are fixed \cite{BMY,brown}. 
For a given instanton action  
functional only ``paths'' that meet these criteria are considered in the full   
path integral. 
For the present calculation we must add boundary terms to the action  
$I = \frac{-1}{16\pi}\int d^4x\sqrt{g}(R-2\Lambda- 
F_{\mu\nu}F^{\mu\nu})$  
that fix both the angular momentum and electric charge of our space-times  
(no additional boundary term is required to fix the magnetic charge).  
A straightforward but tedious calculation of the action for each class 
yields
\begin{equation}
I_{lw,c,n} = -\frac{1}{8}\sum \cA_{lw,c,n}
\end{equation}
which is the sum of the areas of the non-degenerate horizons in each of the lukewarm (lw),
cold (c) and Nariai (n) cases.
Since the pair-creation probability is proportional to $e^{-2I}$,  
the creation rate of these space-times relative to that of   
a deSitter space with the same cosmological constant is $e^{2I_{dS}-2I}$. 
Pure deSitter space is always more likely to be created than a black
hole spacetime as is illustrated in figure 1 which plots $I/I_{dS}$ for   
the instantons.  
\begin{figure}  
\centerline{\psfig{figure=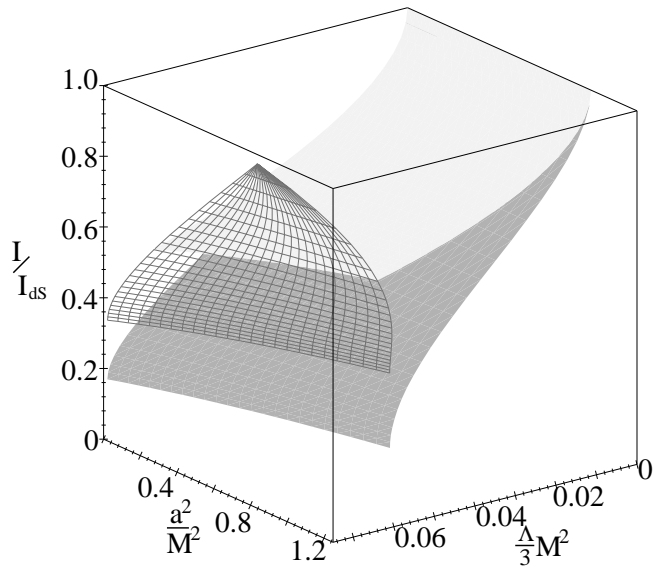,height=8cm}}   
\caption{The actions for the charged and rotating lukewarm, cold, and  
Nariai instantons. The instantons are parameterized by $\frac{a^2}{M^2}$  
and $\frac{\Lambda}{3}M^2$. Their action is plotted as a fraction of  
the action $I_{dS}$ which is the action of the instanton creating pure   
deSitter space with the same  
cosmological constant. The Nariai instantons are the meshed sheet, the  
lukewarm instantons are the light grey sheet, and the cold instantons are  
the dark grey sheet.}  
\end{figure}  
A careful analysis indicates that, for a given value of the mass, 
as the rotation parameter increases (and the charge correspondingly
decreases) the creation rate of lukewarm holes is slightly
enhanced, whereas for the cold and Nariai cases it is slightly
suppressed. Note also that the creation rate for cold black holes 
is suppressed  relative to the non-extreme (lukewarm) black holes by a factor  
of $\exp{\cA_h}$ (where $\cA_h$ is the outer horizon), 
as in the non-rotating cases \cite{dgkt,cosern,strpair}.

In cases where we may regard a space-time as  
being in thermodynamic equilibrium we may  
reinterpret the path integral as a partition function and thereby  
use the instantons to calculate its entropy. For each case
a direct calculation yields $S = - 2I$, and so the   
predicted entropies of the space-times are equal to one-quarter the sum of the 
areas of the  non-degenerate horizons, in agreement with non-rotating cases
\cite{dgkt,cosern,strpair}.

Our approach is contrary to that taken in earlier studies \cite{dp,LP,MM,WZ} 
in which real Euclidean instantons are obtained by
complexifying physical parameters, although the possible 
physical relevance of complex instantons was briefly considered in ref.\ \cite{MM}.
As noted previously \cite{BMY}, such metrics  
have little to do with physical black holes. Indeed, using such a prescription
(e.g. $a\to ia$), 
the fields $h_{ij}$, $E^i$ and $B^j$ induced by the instanton solution 
will no longer match their Lorentzian counterparts on $\Sigma$; furthermore
the root structure  of the function $\cQ$ is altered, 
implying that certain KNdS solutions have corresponding real instantons that  
do not even have the correct number of roots necessary to close them  
up in the manner discussed above. Consequently
it is inconsistent to simultaneously demand that instantons be real and   
Euclidean whilst matching onto their Lorentzian counterparts.   
  
In order for the relationship between   
the entropy of a black hole and its rate of pair creation  
to hold, the black hole  
space-time should be at least quasi-static.   
However, the complex instantons discussed here  
mediate the creation of space-times that are in thermal, but not in  
full thermodynamic equilibrium, as they are unstable to both discharge
and spin-down effects, where the former is expected to 
occur more quickly \cite{page}. However for black holes whose mass is large  
relative to their charge and rotation, discharge and spin  
down  will occur relatively slowly and so the space-time may be  
considered quasi-static even if it is not in full thermodynamic  
equilibrium.  These issues are discussed in more detail in \cite{me}.

We have argued that in order to successfully discuss
stationary-axisymmetric   
space-times within the path integral formulation of  
quantum gravity we must either give up the idea that instantons must  
match onto the Lorentzian space-times that they are creating along a  
hypersurface or we must allow the existence of complex instantons.  
Although this work has been carried out in a cosmological  
background, we expect that these results will apply for any background   
which can  
produce rotating black hole pairs.  Since the matching conditions are  
the only criteria that definitively link a given solution with a  
given instanton it would seem to be essential to include complex  
instantons in  
the generic description of black hole pair-production processes. We  
have seen that complex instantons smoothly match onto Lorentzian  
solutions according to the standard laws for matching solutions to  
the Einstein-Maxwell equations. The pair-creation rates are  
consistent with the interpretation of the entropy as the logarithm  
of the number of quantum states of the black hole.   
  
We conclude that the pair creation of  
all standard black holes may reasonably be described using  
the path-integral formulation of quantum gravity and the no-boundary  
proposal.  
  
\section*{Acknowledgments}  
  
This work was supported by the Natural Sciences and Engineering   
Research  
Council of Canada.  
  
%\section{References}  
  
\newcommand{\PR}{Phys. Rev.}  
\newcommand{\PRL}{Phys. Rev. Lett.}  
\newcommand{\CQG}{Class. Quantum Grav.}  
\newcommand{\CMP}{Commun. Math. Phys.}  
\newcommand{\NP}{Nucl. Phys.}  
\newcommand{\PL}{Phys. Lett.}

\end{document}